\documentclass[a4paper]{jpconf}
\usepackage{graphicx}
\usepackage{cite}
\usepackage{amssymb}

\newcommand{\Br}{\mathrm{Br}}

\newcommand{\cP}{\mathcal{P}}

\def\be{\begin{equation}}
\def\ee{\end{equation}}
\def\beqn{\begin{eqnarray}}
\def\eeqn{\end{eqnarray}}

\def\ba{\begin{array}{c}}
\def\bat{\begin{array}{cc}}
\def\ea{\end{array}}
\def\bi{\begin{itemize}}
\def\ei{\end{itemize}}

\def\cL{{\cal L}}

\usepackage{color}

\definecolor{pink}{rgb}{1,0.5,0.75}
\definecolor{violet}{rgb}{0.5,0,1}

\begin{document}
\title{ \bf $B\to D^{(*)}\tau\nu_\tau$ decays in two-Higgs-doublet models}

\author{Alejandro Celis$^1$, Martin Jung$^2$, Xin-Qiang Li$^{3,1}$, Antonio Pich$^1$ }

\address{$^1$ IFIC, Universitat de Val\`encia -- CSIC, Apt. Correus 22085, E-46071 Val\`encia, Spain}

\address{\small $^2$ Institut f\" ur Physik, Technische Universit\" at Dortmund, D-44221 Dortmund, Germany}
\address{\small $^3$ Physics Department,   
Henan Normal University, Xinxiang, Henan 453007, P.~R. China}

\ead{alejandro.celis@ific.uv.es, martin2.jung@tu-dortmund.de, xqli@itp.ac.cn, antonio.pich@ific.uv.es}

\begin{abstract}
A sizable excess with respect to the SM expectation has been reported recently by the BaBar collaboration in the decay rates $B  \rightarrow D^{(*)} \tau \nu$,
normalized by the corresponding light lepton modes.   
A violation of lepton flavor universality as suggested by this excess could be due to a charged Higgs mediating these processes at tree level.   In this talk we analyze the implications of the observed excess 
within the framework of two-Higgs-doublet models, considering also the bounds from other semileptonic and leptonic decays of $B$ and $D_{(s)}$ mesons.  Prospects for  
$ B  \rightarrow D^{(*)} \tau \nu$ decays at future Super-Flavor Factories are also discussed.
\end{abstract}

\section{Introduction}
The hypothesis that a scalar sector in nature is behind the origin of particle masses (or the mechanism of spontaneous electroweak symmetry breaking) is starting to look even more plausible since the discovery of a new boson by the ATLAS and CMS collaborations with characteristics resembling those of the Standard Model (SM) Higgs boson~\cite{:2012gk,:2012gu}.   The SM scalar sector is composed of one doublet of complex scalar fields, containing in total four real degrees of freedom; three of them
give mass to the $W^{\pm}$ and $Z$ bosons, while the remaining real field becomes a physical scalar, the Higgs boson.  In the SM the fermions also acquire their masses from the spontaneous breaking of the electroweak symmetry due to their interactions with the Higgs. 

There is nothing fundamental a priori, however, preventing the scalar sector to have a richer structure, for example that of a Two-Higgs-Doublet Model (2HDM), where
two doublets of complex scalar fields are responsible for the breaking of the electroweak symmetry.     The physical scalar spectrum includes
in this case three neutral bosons and one charged scalar particle.    The boson discovered at the LHC
could be in principle any of the three neutral bosons of a 2HDM.

The BaBar collaboration has recently reported an excess of events in two semileptonic transitions of the type $b\to c\,\tau  \nu_\tau$. More specifically, they have measured the ratios~\cite{Lees:2012xj}
\beqn\label{eq:Babar}
R(D)& \equiv &\frac{\Br(\bar B\to D\tau^-\bar\nu_\tau)}{\Br(\bar B\to D\ell^-\bar\nu_\ell)}\; \stackrel{\rm BaBar}{=}\; 0.440\pm 0.058\pm 0.042\;\stackrel{\rm avg.}{=}\; 0.438\pm0.056\, ,
\nonumber\\[5pt]
R(D^*)& \equiv &\frac{\Br(\bar B\to D^*\tau^-\bar\nu_\tau)}{\Br(\bar B\to D^*\ell^-\bar\nu_\ell)}\; \stackrel{\rm BaBar}{=}\; 0.332\pm 0.024\pm 0.018\;\stackrel{\rm avg.}{=}\; 0.354\pm0.026\, ,
\eeqn
which are normalized to the corresponding decays into light leptons $\ell=e,\mu$.    The second value given in each line is the average with the previous measurements by the Belle collaboration~\cite{Adachi:2009qg,Bozek:2010xy}.        The decays $B \rightarrow D^{(*)} \tau \nu$ involve the heaviest fermions that can be directly produced at flavor factories and constitute a well-suited place to look for new physics effects related to the mechanism of electroweak symmetry breaking.    A charged Higgs arising from a 2HDM can in principle cause significant deviations from the SM in semileptonic and leptonic decays where it enters at tree level.   Moreover, a charged Higgs leads  naturally to violations of lepton flavor universality due to the characteristic mass dependence of its couplings to fermions.    Interestingly, a charged Higgs boson could also produce an enhancement in the di-photon decay of the neutral Higgs, as currently observed by the ATLAS and CMS collaborations~\cite{:2012gk,:2012gu}.

In this talk we are mainly interested in addressing two questions:
\begin{itemize}
\item[{\bf A)}] What is the required flavor structure for a 2HDM in order to accommodate the observed excess in $R(D^{(*)})$, as well as other measured semileptonic and leptonic decays of $B$ and $D_{(s)}$ mesons?

\item[{\bf B)}] How can we discriminate between different new physics scenarios that can explain the excess in $R(D)$ and $R(D^*)$?
\end{itemize}

To answer the first question, we analyze in Section~\ref{sec1} the bounds on the charged Higgs couplings from different semileptonic and leptonic decays where the charged Higgs contribution enters at tree level.
The second question is considered in Section~\ref{sec2}, where we study the effect of charged scalar contributions in new observables involving angular distributions, polarizations of the final $\tau$ and $D^*$ meson, as well as the momentum-transfer dependence.   Our discussion will be focused on results presented in~\cite{Jung:2010ik,Celis:2012dk}. These questions have also been addressed recently by different authors~\cite{Nierste:2008qe,Kamenik:2008tj,Tanaka:2010se,Fajfer:2012vx,Sakaki:2012ft,Fajfer:2012jt,Datta:2012qk,Crivellin:2012ye,Tanaka:2012nw}.

\section{Explaining the excess within the 2HDM}
\label{sec1}

Four-fermion interactions mediated by a charged scalar are described by the Lagrangian
\be\label{genlagrangian}
\cL_{\rm eff} \; =\; -\frac{4 G_F}{\sqrt{2}}\;\sum_{q=u,c}\, V_{qb}\;\sum_{l=e,\mu,\tau}\;\left\{
\left[\bar q\gamma^\mu \cP_L b\right] \left[\bar l\gamma_\mu\cP_L\nu_l\right]
\, +\, \left[\bar q \left( g_L^{qbl}\,\cP_L + g_R^{qbl}\, \cP_R\right) b\right] \left[\bar l\cP_L\nu_l\right]
\right\}\, ,
\ee
as long as the charged Higgs is heavy enough to be integrated out.  This is a very good approximation, given that direct searches at LEP for charged Higgs bosons have set the lower limit $m_{H^{\pm}}  \gtrsim 80$~GeV~\cite{Searches:2001ac},  assuming that it couples to fermions.   In Equation (\ref{genlagrangian}), $V_{qb}$ refers to the corresponding element of the CKM mixing matrix, $G_F$ is the Fermi constant and $ \cP_{L, R}$ are the left- and right-handed chiral projectors, respectively.  The couplings of the charged Higgs to fermions are contained in the dimensionless parameters $g_L^{qbl}$ and $g_R^{qbl}$, which   
depend on the assumed Yukawa structure of the 2HDM.   In the following we will treat them as independent complex quantities to keep the discussion as general as possible.   Explicit expressions for $g_L^{qbl}$ and $g_R^{qbl}$ in terms of the 
general parametrization of the Aligned 2HDM (A2HDM)~\cite{Pich:2009sp}
are provided in~\cite{Celis:2012dk}.   The different types of 2HDMs based on discrete ${\cal Z}_2$ symmetries can be obtained as particular limits of this parametrization.

\begin{figure}[thb]
\centering
\includegraphics[height=5.2cm]{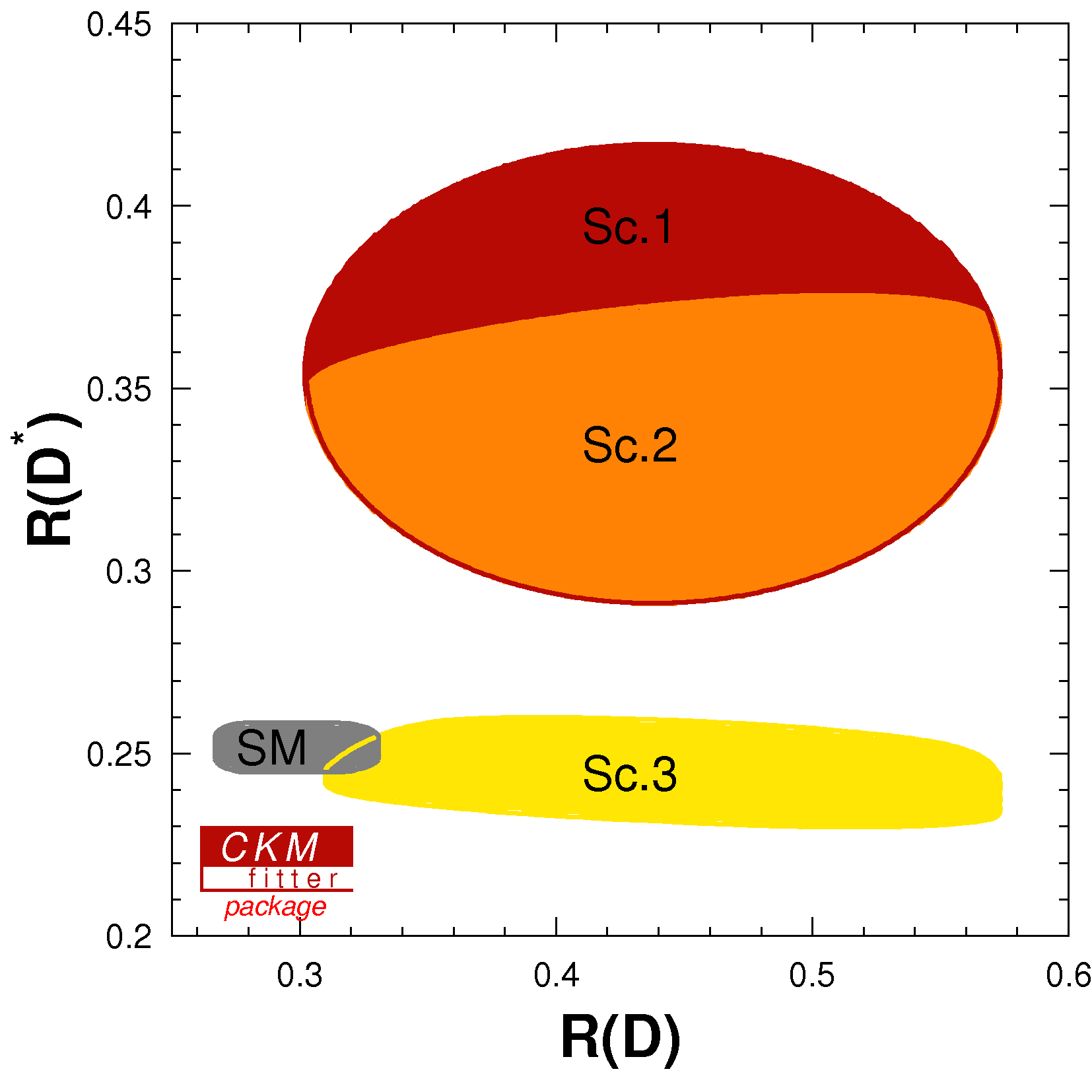}
\includegraphics[height=5.2cm]{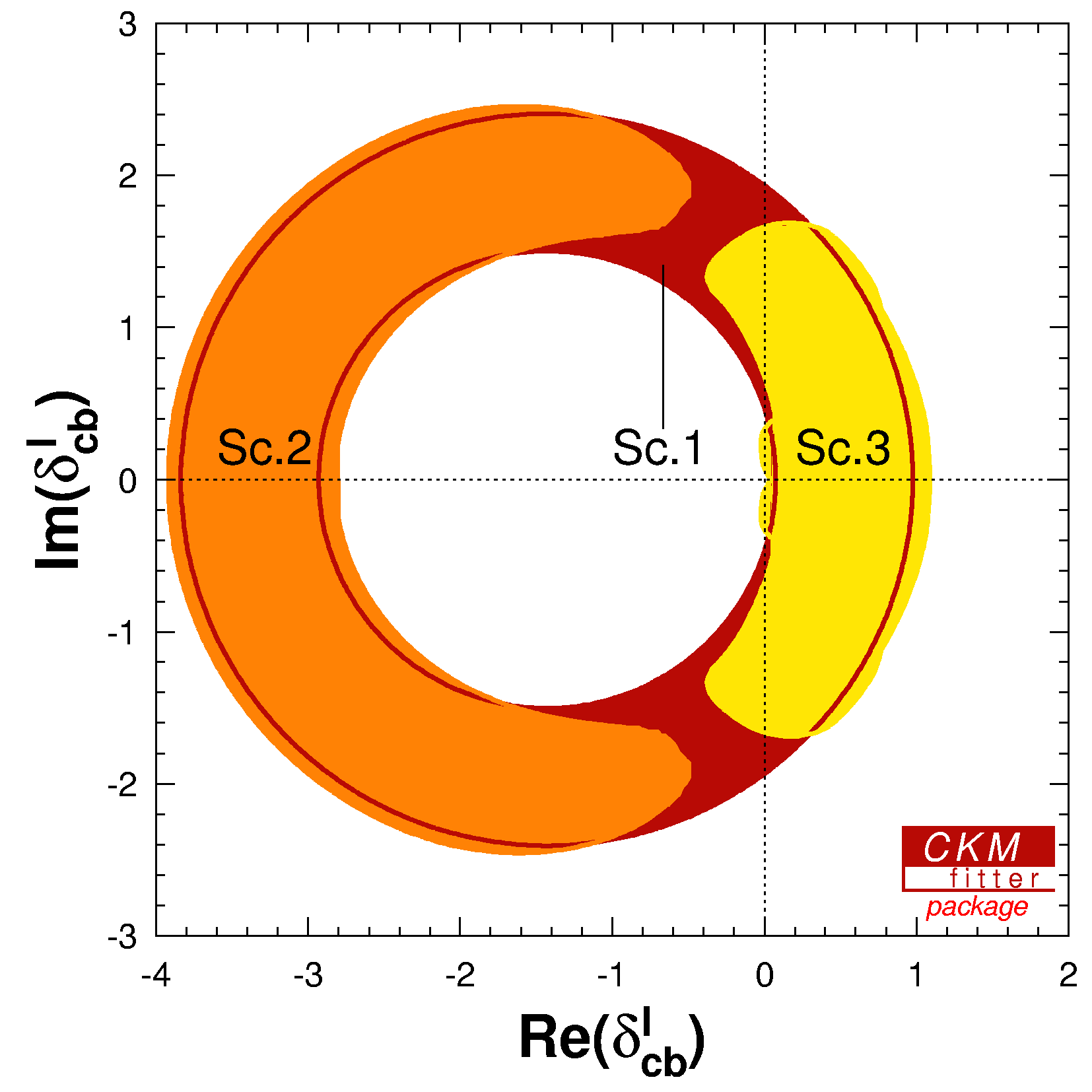}
\includegraphics[height=5.2cm]{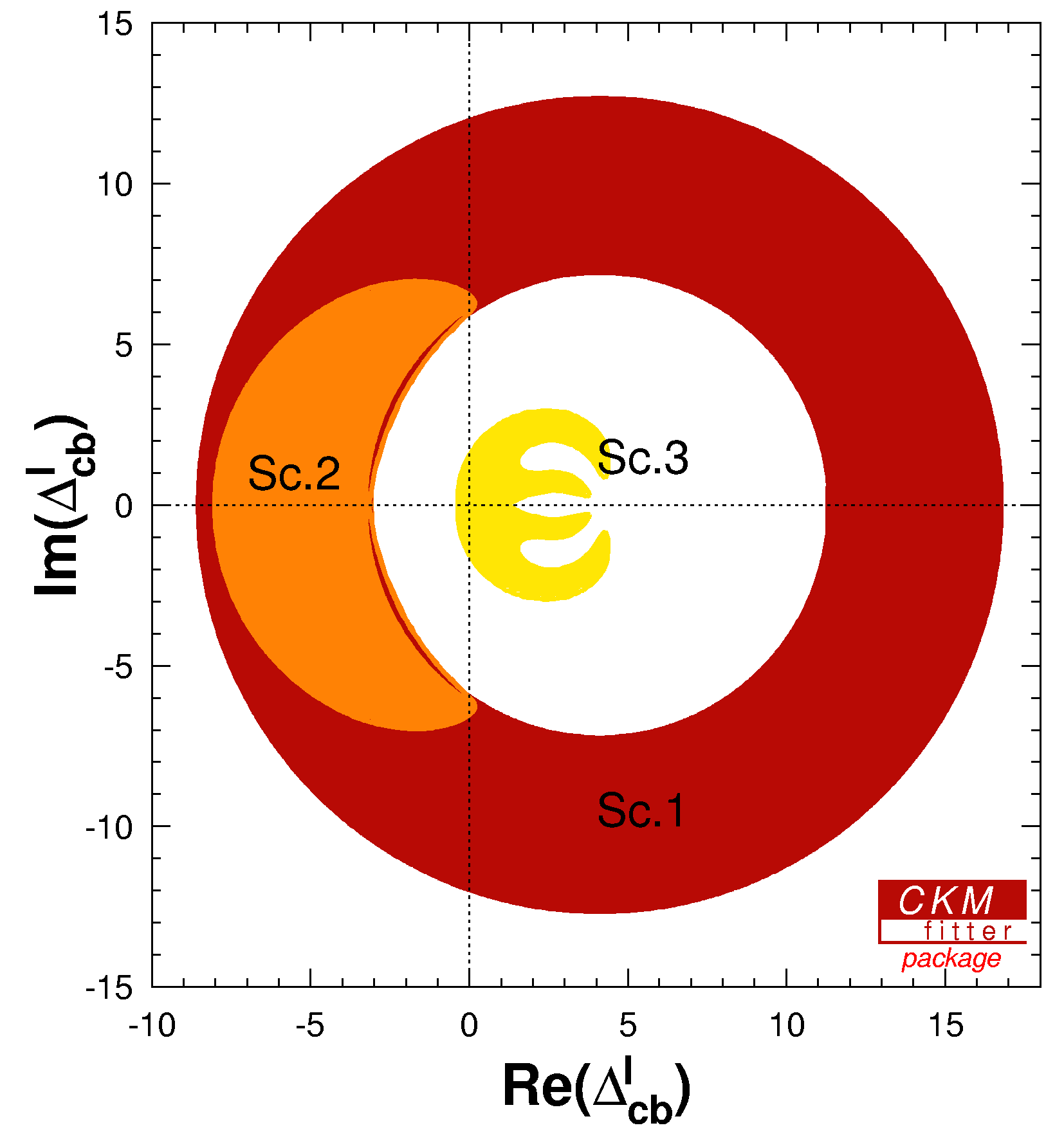}
\caption{\label{zetadlul-allowed} \it {\small Allowed regions in the $R(D)$--$R(D^*)$~(left), complex $\delta_{cb}^l$~(center) and} $\Delta_{cb}^l$~(right) planes at $95\%$~CL, corresponding to the three different scenarios. See text for details.}
\end{figure}

Due to the different spins of the $D$ and $D^{*}$ mesons, the parameters $g_L^{qbl}$ and $g_R^{qbl}$ appear in different linear combinations in $R(D)$ and $R(D^*)$,
\begin{equation}  \label{eq::Delta} 
\delta_{cb}^l\,\equiv\, \frac{(g_L^{cbl}+g_R^{cbl})(m_B-m_D)^2}{m_l(\overline{m}_b-\overline{m}_c)}\,,
\qquad\qquad
\Delta_{qb}^l\, = \,\frac{(g_L^{qbl}-g_R^{qbl}) m_B^2}{m_{l}(\overline{m}_{b} + \overline{m}_q)}\; \,.
\end{equation}
%
$R(D)$ depends only on the scalar combination $\delta_{cb}^l$ while $R(D^*)$ only depends on the pseudo-scalar one $\Delta_{qb}^l$~\cite{Celis:2012dk}.   In versions of the 2HDM were the fermionic charged Higgs couplings are proportional to fermion masses and the proportionality factor is the same for all families (family universality), the following relations hold:
%
%
\begin{equation} \label{eq::gratios}
\frac{g_L^{q_{u}q_{d}l}}{g_L^{q'_{u}q'_{d}l'}}\, =\, \frac{m_{q_u}m_l}{m_{q'_u}m_{l'}}  \,,
\qquad\qquad\qquad
\frac{g_R^{q_{u}q_{d}l}}{g_R^{q'_{u}q'_{d}l'}}\, =\, \frac{m_{q_d}m_l}{m_{q'_d}m_{l'}} \,.
\end{equation}
The Aligned Two-Higgs doublet model as well as the different types of 2HDMs based on discrete ${\cal{Z}}_2$ symmetries (Types I, II, X and Y)  belong to this class of models~\cite{Branco:2011iw}.   To analyze possible deviations of the hypothesis of family universality in the context of 2HDMs we consider three scenarios with different flavor structures:
\begin{enumerate}
\item[$\bullet$] Scenario~1:  
The Yukawa couplings $g_{L,R}^{q_{u}q_{d}l}$ are assumed to be independent for the
different transitions ($b \rightarrow u,c\, ;\, c \rightarrow d\, \ldots$).
We therefore fit only $R(D)$ and  $R(D^*)$.
\item[$\bullet$] Scenario~2:  We assume that the hypothesis of family universality (\ref{eq::gratios}) holds for $q_d= b$ and any up-type quark.  The charged scalar contribution to $b \rightarrow c$ and $b \rightarrow u$ transitions are then related.  In this case we consider the bounds from $R(D^{(*)})$ and $\Br(B \rightarrow \tau \nu)$.
\item[$\bullet$] Scenario~3:  The hypothesis of family universality (\ref{eq::gratios}) is assumed to hold for all quark generations.     In this case we consider in the fit all observables in Table \ref{tab::SM} except for $R(D^*)$.   If $R(D^*)$ is included in the fit no allowed region is found at  $95\%$ CL.
\end{enumerate}
Experimental values and SM predictions for the different observables used are listed in Table~\ref{tab::SM}~\cite{Celis:2012dk}.  In Figure~\ref{zetadlul-allowed} we show $95\%$~CL allowed regions in the parameters $\delta_{cb}^l$ and $\Delta_{cb}^l$ for the three scenarios. 
 We can now give an answer to question A.

\begin{table}[tb]
\begin{center}
\caption{\label{tab::SM} \it \small Predictions within the SM for the various semileptonic and leptonic decays considered in this work, together with their corresponding experimental values. The first uncertainty given always corresponds to the statistical error, and the second, when given, to the theoretical one.}
\vspace{0.2cm}
\doublerulesep 0.8pt \tabcolsep 0.07in
\small{
\begin{tabular}{lccc}
\hline\hline
Observable   					&  SM Prediction					& Exp. Value  \\
\hline \\[-10pt]
$R({D})$ 							& $0.296^{+0.008}_{-0.006}\pm0.015$ & $ 0.438 \pm 0.056$ \\
$R({D^*})$       					& $0.252\pm0.002\pm0.003$ 		& $0.354 \pm 0.026$					 \\
$\Br(B\to \tau \nu_\tau)$ 	&  $(0.79^{+0.06}_{-0.04}\pm0.08)\times 10^{-4}$& $(1.15 \pm 0.23)\times 10^{-4}$	 \\
$\Br(D_s \to \tau \nu_\tau)$	&  $(5.18 \pm 0.08\pm0.17) \times 10^{-2} $& $(5.54 \pm 0.24)\times 10^{-2}$  \\
$\Br(D_s \to \mu \nu)$	& $(5.31 \pm 0.09\pm0.17) \times 10^{-3} $	& $(5.54 \pm 0.24)\times 10^{-3}$ \\
$\Br(D \to \mu \nu)$	&  $(4.11^{+0.06}_{-0.05}\pm0.27) \times 10^{-4} $ & $(3.76 \pm 0.18)\times 10^{-4}$	 \\
$\Gamma(K\to\mu\nu)/\Gamma(\pi\to\mu\nu)$    & $1.333\pm0.004\pm0.026$ & $1.337\pm0.003$ \\
$\Gamma(\tau\to K\nu_\tau)/\Gamma(\tau\to\pi\nu_\tau)$ & $(6.56\pm0.02\pm0.15)\times10^{-2}$ & $(6.46\pm0.10)\times10^{-2}$ \\
\hline\hline
\end{tabular}}
\end{center}
\end{table}
%

All $B$ decays, $R(D^{(*)})$ and $B \rightarrow \tau \nu_{\tau}$, can be accommodated within a general 2HDM with family universality for $q_d =b$ and any  up-type quark; i.e. in scenarios 1 and 2.  However this is not possible in the usual Type II 2HDM; 
a more general flavor structure
is needed in which the charged scalar couplings to up-type quarks are not as suppressed~\cite{Celis:2012dk}.

It is not possible to accommodate all the observables in Table~\ref{tab::SM} simultaneously in a 2HDM with family universality between all fermion generations (scenario 3).  Since charged scalar contributions to $R(D^*)$ are subdominant compared with the vector contributions, large Yukawa
couplings are required to explain the observed excess in $R(D^*)$, thus generating a tension with present bounds from $D_{(s)}$ leptonic decays.
If $R(D^*)$ is not considered in the fit, all observables can be accommodated within the family universal A2HDM.   Should the
  present measurement be confirmed, a departure from the hypothesis of flavor universality is therefore needed in order to account for all semileptonic and leptonic decays of $B$ and $D_{(s)}$ mesons. 
Bounds from loop-induced processes discussed in \cite{Jung:2010ik}, also point towards a violation of family universality in the coupling with the top quark, once the present excess in $R(D^{(*)})$ is considered~\cite{Celis:2012dk}.

\section{Sensitivity of new observables to charged scalar contributions}
\label{sec2}

Using only measurements of branching fractions, it is not possible to disentangle
charged scalar contributions in $B \rightarrow D^{(*)} \tau \nu$ from other kinds of new physics.  However,
$B \rightarrow D^{(*)} \tau \nu$ decays have a rich three-body kinematics and spin structure in the final state that has not been exploited so far.    Several observables involving angular distributions, polarization fractions and momentum-transfer dependence,  would provide, if measured, crucial information to discriminate between different new physics scenarios and clarify the possible role of charged scalar contributions in these processes~\cite{Korner:1989qb,Hagiwara:1989gza,Tanaka:1994ay,Chen:2005gr,Chen:2006nua,
Nierste:2008qe,Tanaka:2010se,Fajfer:2012vx,Sakaki:2012ft,Datta:2012qk}.  

One way to test the observed excess in $R(D^*)$ would be to measure the leptonic decay $B_c \rightarrow \tau \nu$, which has the same flavor structure $b \rightarrow c \tau \nu_{\tau}$ and is sensitive to the same combination of Yukawa couplings.
Including the charged scalar contribution, the decay width is given by
%
\begin{equation}\label{eq:Gamma_Plnu}
\Gamma(B_c \to\tau \nu_\tau)\, =\, G_F^2m_\tau^2f_{B_c}^2|V_{cb}|^2 \,\frac{m_{{B_c}}}{8\pi} \left( 1- \frac{m_\tau^2}{m_{B_c}^2} \right)^2  \; |1-\Delta^\tau_{cb}|^2 \; ,
\end{equation}	
where $f_{B_c}$ is the $B_c$ leptonic decay constant.
Recent Lattice QCD calculations find $f_{B_c}  = 0.427 (6)(2)$~ GeV, giving the SM prediction $\Br(B_c\to\tau\nu) = 0.0194 (18)$~\cite{12070994}.  Large enhancements of $\Br(B_c\to\tau\nu)$ should be observed in principle if the current excess in $R(D^*)$ is due to charged scalar contributions.

\begin{figure}[thb]
\centering
\includegraphics[width=7.8cm,height=5.0cm]{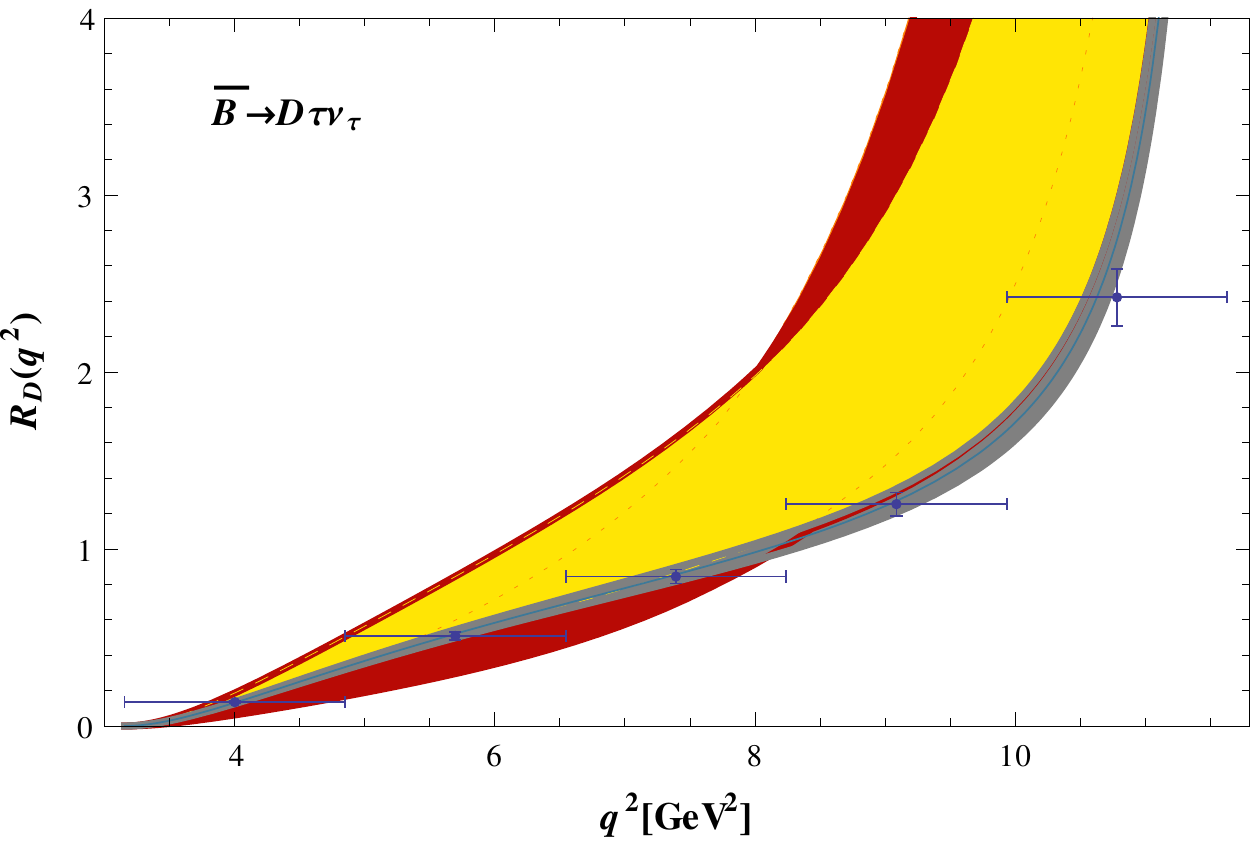}
~
\includegraphics[width=7.8cm,height=5.0cm]{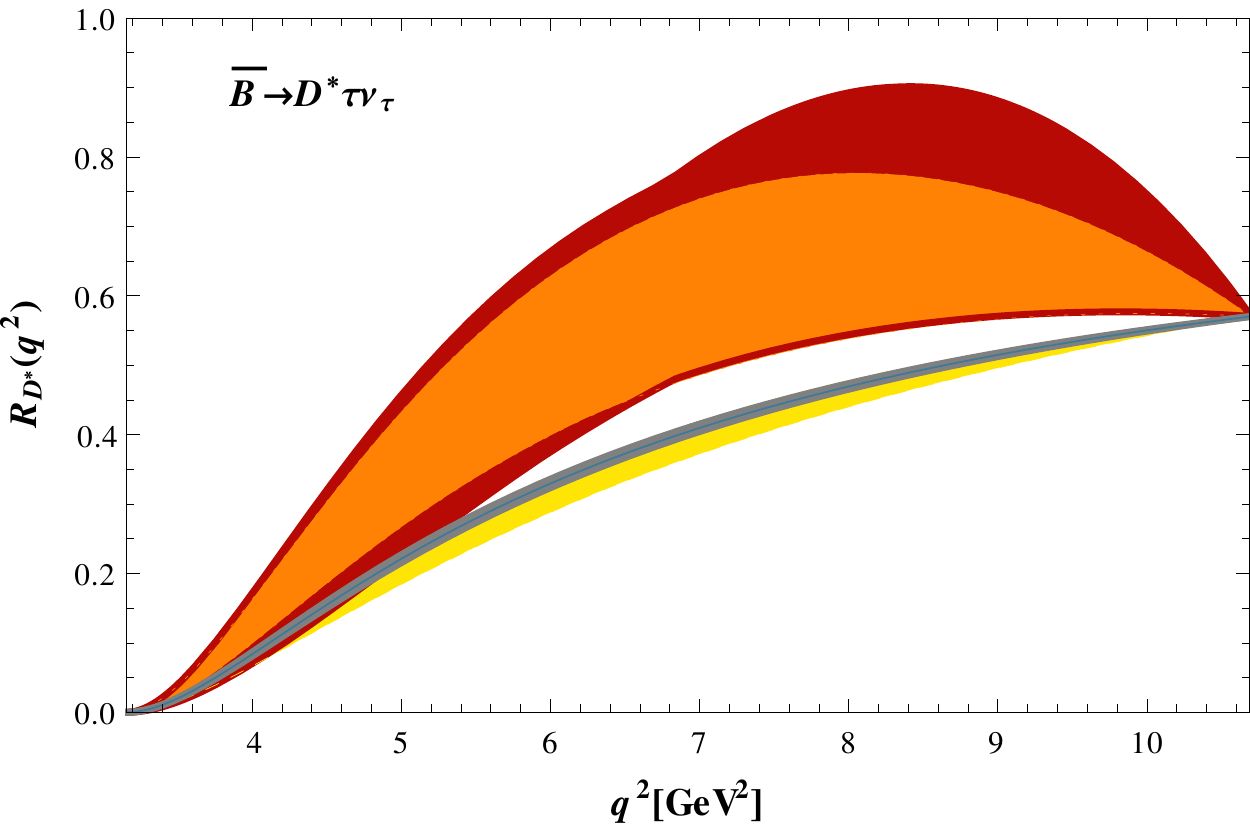}
\caption{\label{fig:dBrs-DRs} \it \small The $q^2$ dependence of the ratios $R_{D^{(*)}}(q^2)$, both within the SM~(gray) as well as in scenarios~1~(red), 2~(orange), and 3~(yellow). The $95\%$ CL allowed regions are obtained with the corresponding experimental constraints exerted for the three different scenarios, and with the hadronic uncertainties added in quadrature for the SM. The five-binned distribution for $R_D(q^2)$ in the SM is also shown. }
\end{figure}

\begin{figure}[tbh]
\centering
\includegraphics[height=5.8cm,width=5.9cm]{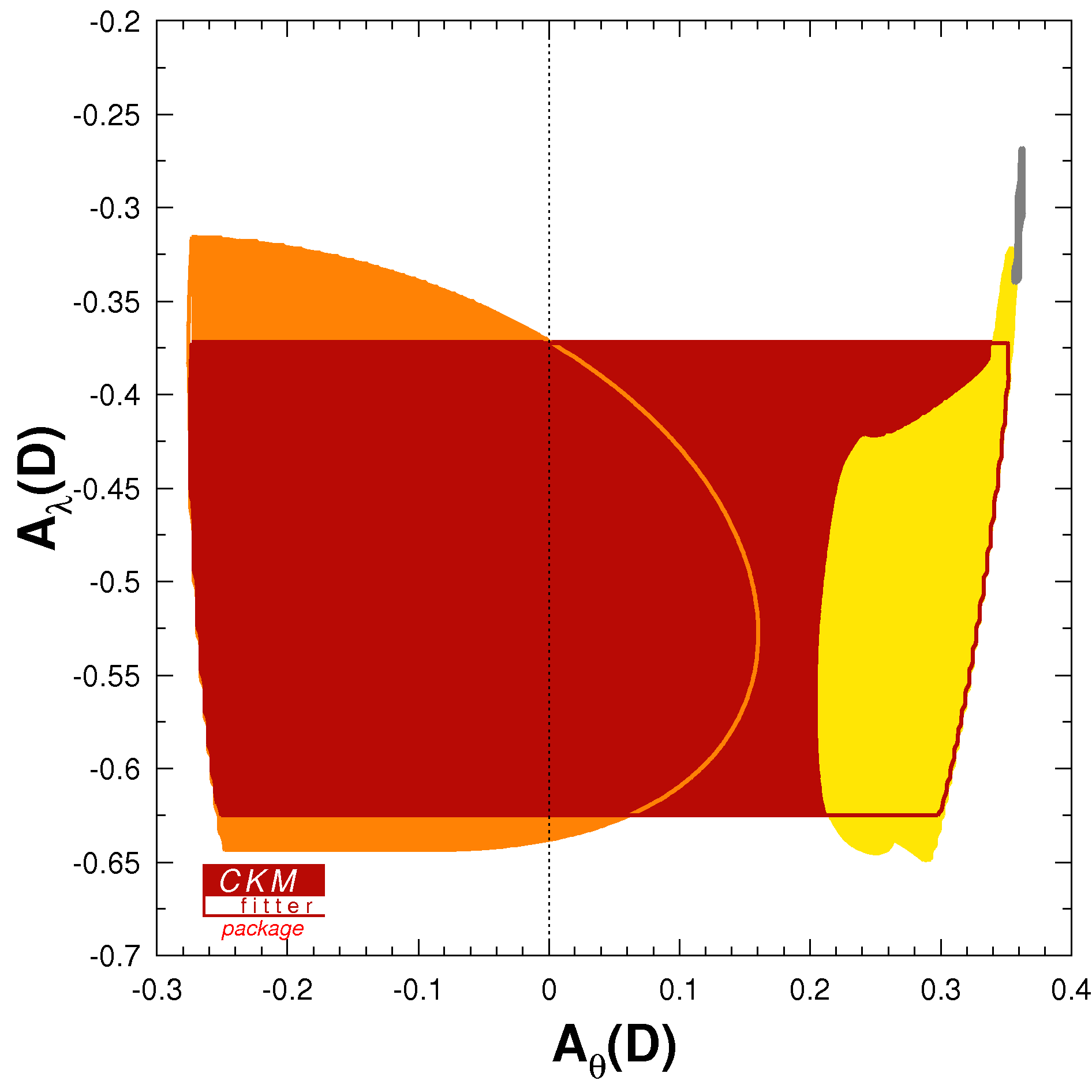}
\includegraphics[height=5.8cm,width=5.9cm]{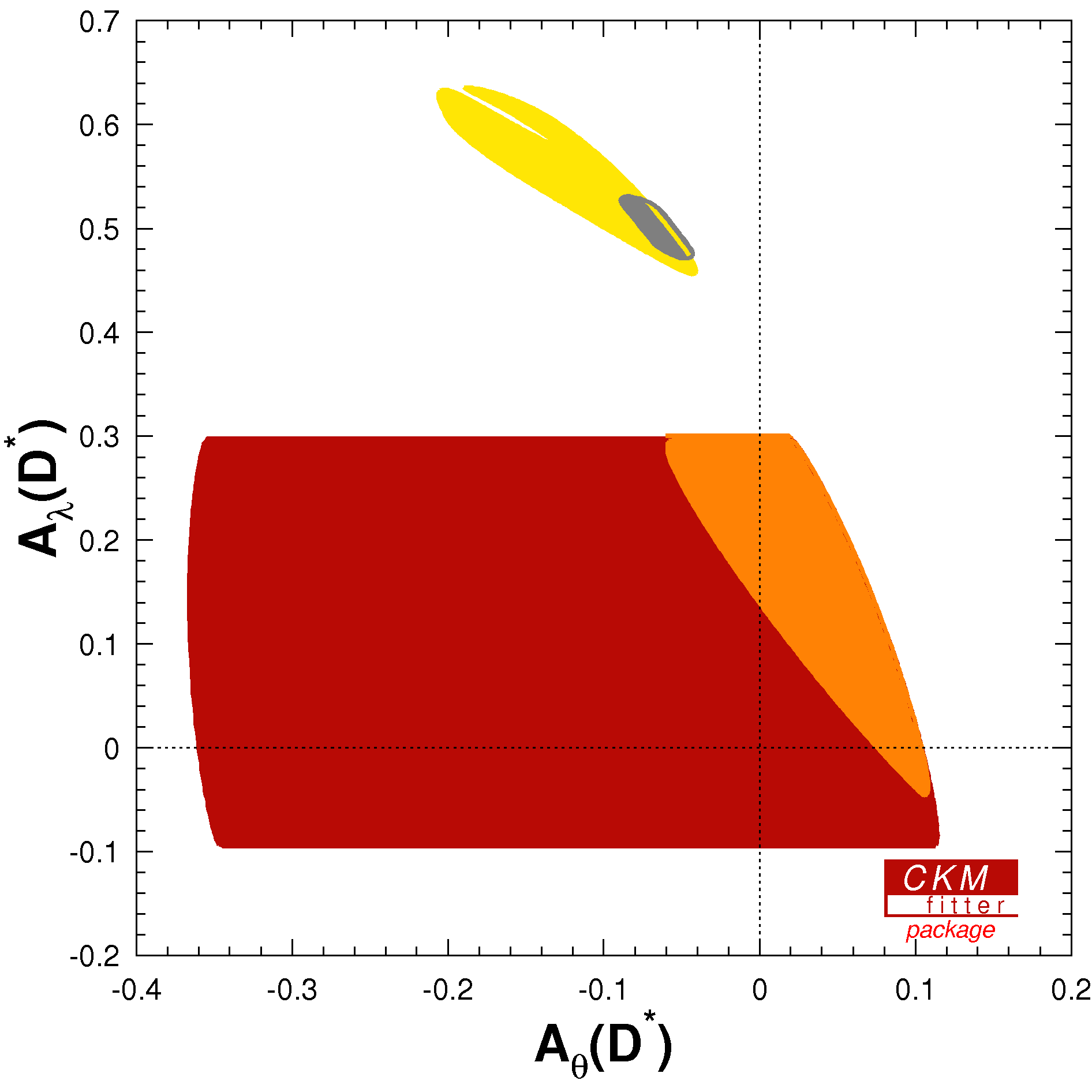}
\includegraphics[width=7.2cm,height=5.4cm]{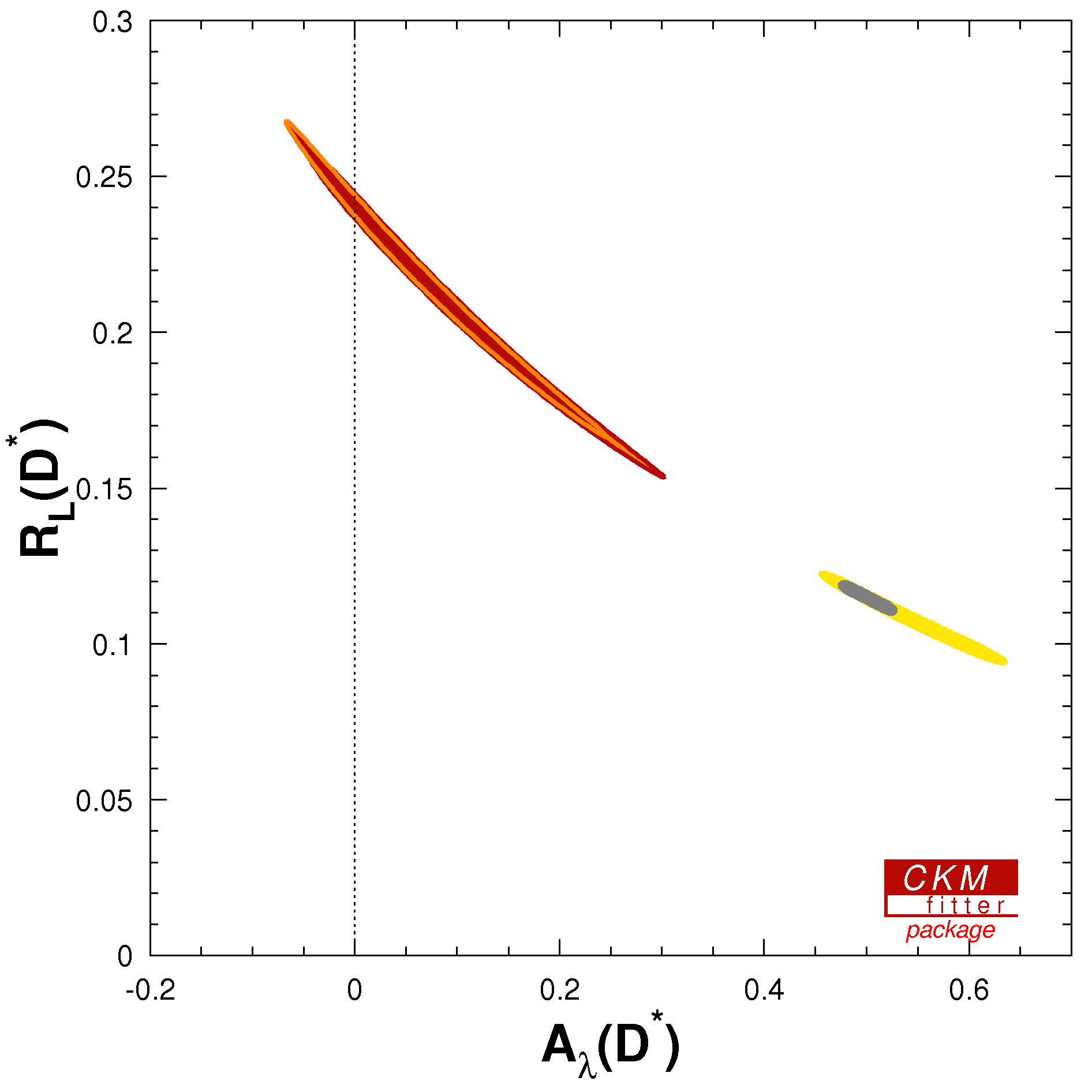}
\caption{\label{fig::AFBAlambda} \it \small In the upper panel we show the prediction for $A_{\theta}(D^{(*)})$ vs. $A_{\lambda}(D^{(*)})$ for the SM (gray), and the three scenarios (1-red, 2-orange, 3-yellow), from a global fit including all appropriate observables. In the lower panel we show the predictions for $A_{\lambda}(D^*)$ vs. $R_L(D^*)$.   }
\centering
\end{figure}

Information about the momentum-transfer
distributions of $R(D)$ and $R(D^*)$,
\begin{equation} \label{eq:dRs}
R_{D^{(*)}}(q^2)\, =\, \frac{d\Br(\bar B\to D^{(*)} \tau^-\bar\nu_\tau)/d q^2}{d\Br(\bar B\to D^{(*)} \ell^-\bar\nu_{\ell})/d q^2} \,,
\end{equation}
would be very useful to determine what kind of new physics could be responsible for the observed excess.
In Figure~\ref{fig:dBrs-DRs} we show the predictions for the $q^2$ distributions in the SM and the three scenarios considered in the previous section.
 Given that charged scalars do not not affect helicity amplitudes with transversely polarized $D^*$ mesons,  the measurement of these distributions for longitudinally polarized $D^*$ may provide a better sensitivity to scalar effects
\begin{equation}
R_L^*(q^2)\, =\,  \frac{d \Gamma^{L}_{\tau}/d q^2}{d \Gamma^{L}_{\ell}/d q^2} \,.
\end{equation}
A charged Higgs mediating $B \rightarrow D^{(*)} \tau \nu$ decays would also have an effect on the polarization of the final $\tau$ lepton.      In the SM, the $W^-$ couples exclusively to left-handed $\tau^-$ while a charged Higgs $H^-$ couples to right-handed $\tau^-$.      Due to their short lifetime, $\tau$ leptons decay within the detector and their polarization can be determined by analyzing the kinematics of the decay products.     One could measure a $\tau$ spin asymmetry defined in the $\tau$-$\bar\nu_{\tau}$ center-of-mass frame as
\begin{equation} \label{eq:ATAUD}
A^{D^{(*)}}_{\lambda}(q^2)\, =\, \frac{ d\Gamma^{D^{(*)}}[\lambda_{\tau} =  -1/2]/ d q^2  - d\Gamma^{D^{(*)}}[\lambda_{\tau} = + 1/2] /d q^2 }{ d\Gamma^{D^{(*)}}[\lambda_{\tau} =  -1/2]/ d q^2  + d\Gamma^{D^{(*)}}[\lambda_{\tau} = + 1/2] /d q^2}\,.
\end{equation}
Angular distributions are also sensitive to the Dirac structure of possible new physics mediating $b \rightarrow c \tau \nu$ transitions.  
A useful observable is the forward-backward asymmetry,  defined as the difference of partial decay rates with the angle $\theta$ between the $D^{(*)}$ and $\tau$ three-momenta in the $\tau$-$\bar\nu_{\tau}$ center-of-mass frame greater or smaller than $\pi/2$:
\begin{equation} \label{eq:AFBDs}
A^{D^{(*)}}_{\theta}(q^2)\, =\, \frac{ \int_{-1}^{0} d \cos\theta\, ( d^2 \Gamma^{D^{(*)}}_{\tau}/dq^2 d \cos\theta )  -\int_{0}^{1} d \cos\theta\, ( d^2 \Gamma^{D^{(*)}}_{\tau}/dq^2 d \cos\theta ) }{ d \Gamma^{D^{(*)}}_{\tau}/dq^2 } \,.
\end{equation}
In Table~\ref{tab::NOSM} we show predictions for the observables discussed previously in the SM as well as in the three scenarios considered.   Correlations between the different observables can be observed in Figure~\ref{fig::AFBAlambda}.

\begin{table}[t]
\begin{center}
\caption{\label{tab::NOSM} \it \small Predictions for the $q^2$-integrated observables both within the SM and in the different scenarios.}
\vspace{0.2cm}
\doublerulesep 0.8pt \tabcolsep 0.04in
\small{
\begin{tabular}{l c c c c}\hline\hline
Observable	        &   SM Prediction                       &   Scenario~1	
                    &   Scenario~2	                        &   Scenario~3 
\\ \hline\\[-10pt]
$ R_L(D^*)$ 	        & $0.115 \pm 0.001 \pm 0.003$           & $0.217\pm0.026$
                    & $0.223^{+0.013}_{-0.026}\pm0.006$	    & $0.104^{+0.006}_{-0.003}\pm0.003$\\

$A_{\lambda}({D})$ 	& $-0.304\pm0.001\pm0.035$				& $-0.55^{+0.10}_{-0.04}$			 
                    & $-0.55^{+0.09}_{-0.04}$			    & $-0.55^{+0.09}_{-0.04}\pm0.01$\\
$A_{\lambda}({D^*})$ & $0.502^{+0.005}_{-0.006}\pm 0.017$    & $0.06^{+0.10}_{-0.06}$			 
                    & $0.04^{+0.10}_{-0.03}\pm0.01$	    	& $0.57^{+0.04}_{-0.02}\pm0.02$\\
$A_{\theta}({D})$ 	& $0.3602^{+0.0006}_{-0.0007}\pm 0.0022$& $0.03^{+0.01}_{-0.00}\pm0.30$		
                    & $-0.21^{+0.13}_{-0.00}\pm0.06$	    & ${0.36^{+0.01}_{-0.09}}^\dagger$\\
$A_{\theta}({D^*})$	& $-0.066\pm 0.006 \pm 0.009 $          & $-0.136^{+0.012}_{-0.003}\pm0.222$
                    & $0.081^{+0.008}_{-0.059}\pm0.009$	    & $-0.146^{+0.039}_{-0.017}\pm0.021$\\
\hline\hline
\end{tabular}
}
\end{center}
\end{table}

%

An unambiguous way to test the hypothesis of a charged scalar mediating $B$ semileptonic decays is to look for observables that are independent of such contributions.    A deviation of these observables from the SM prediction could then only be attributed to new physics different from a charged scalar.    One such observable is the ratio of leptonic $B$ decay rates into $\tau$ and $\mu$,
\begin{equation}  \label{ratioB}
\frac{ \Br(   B \rightarrow \tau \nu )  }{  \Br(  B \rightarrow \mu \nu ) }\, = \, \frac{m^2_{\tau}}{m^2_{\mu}}   \left( \frac{ 1 - m^2_{\tau}/m_{B}^2 }{ 1 - m^2_{\mu}/m_{B}^2   }   \right)^2 \,,
\end{equation}
which is valid for $B_u$ and $B_c$. Some
combinations of observables that are insensitive to charged scalar contributions in $ B \rightarrow D^{(*)} \tau \nu_{\tau}$ transitions were defined in~\cite{Celis:2012dk}:
\begin{equation}\label{eq::X1}
X_1(q^2)\,\equiv\, R_{D^*}(q^2)-R_L^*(q^2)\, ,
\qquad\qquad
X_2^{D^{(*)}}(q^2)\,\equiv\, R_{D^{(*)}}(q^2)\,\left[ A^{D^{(*)}}_{\lambda}(q^2)+1\right]\,.
\end{equation}

\section{Summary}
$B \rightarrow D^{(*)} \tau \nu$ decays are an excellent laboratory to look for new physics effects related to the mechanism of electroweak symmetry breaking.    They involve the heaviest fermions that can be directly produced at flavor factories.    They also possess a rich three-body kinematics and spin structure that could be fully exploited at future Super-Flavor Factories, providing important information about possible new physics mediating these transitions.

In this talk we presented an analysis of charged scalar contributions to these processes and then derived some implications for the flavor structure of 2HDMs in order to explain the excess observed by BaBar~\cite{Lees:2012xj}.  We showed that, in contrast to the 2HDMs with $\mathcal{Z}_2$ symmetry, the A2HDM can explain the present data on $R(D^{(*)})$ and $B \rightarrow \tau \nu$, but that the excess observed in $R(D^*)$ is not compatible with current bounds from leptonic $D_{(s)}$ decays within this framework.  If the
  present excess in $B \rightarrow D^{(*)} \tau \nu$ is confirmed,  it will be necessary to consider a departure from the family universality of the Yukawa couplings in order to 
accommodate all data from $B$ and $D_{(s)}$ decays within the framework of 2HDMs.

New theoretical developments on the calculation of the hadronic matrix elements would improve our knowledge of the SM prediction for these decays and shed light on a possible new-physics interpretation of the observed excess; for recent work on the determination of $B \rightarrow D$ hadronic matrix elements, see~\cite{Becirevic:2012jf,Bailey:2012jg}.  Certainly new Belle measurements are needed to clarify this puzzle.   Preliminary efforts to measure this decays at LHCb have already been done~\cite{John}.    We have also discussed new observables in $b \rightarrow c \tau  \nu$ transitions that could be measured at future Super-Flavor factories and test the presence of charged scalars mediating these decays; these involve angular distributions, polarization of $\tau$ and $D^*$ as well as the $q^2$ dependence.

\ack
 This work has been supported in part by the Spanish Government [grants FPA2007-60323, FPA2011-23778 and CSD2007-00042 (Consolider Project CPAN)]. X.Q.L. is also supported in part by the National Natural Science Foundation of China under contract No. 11005032, the Specialized Research Fund for the Doctoral Program of Higher Education of China (Grant No. 20104104120001) and SRF for ROCS, SEM. M.J. is supported by the Bundesministerium f\"ur Bildung und Forschung.
The work of A.C. is funded through an FPU grant (AP2010-0308, MINECO, Spain).


\section*{References}
\begin{thebibliography}{99}

\bibitem{:2012gk}
  G.~Aad {\it et al.}  [ATLAS Collaboration],
  ``Observation of a new particle in the search for the Standard Model Higgs boson with the ATLAS detector at the LHC,''
  Phys.\ Lett.\ B {\bf 716} (2012) 1
  (\textit{Preprint}  hep-ex/1207.7214)


\bibitem{:2012gu}
  S.~Chatrchyan {\it et al.}  [CMS Collaboration],
  ``Observation of a new boson at a mass of 125 GeV with the CMS experiment at the LHC,''
  Phys.\ Lett.\ B {\bf 716} (2012) 30
  (\textit{Preprint}  hep-ex/1207.7235)


\bibitem{Lees:2012xj}
  J.~P.~Lees {\it et al.}  [BaBar Collaboration],
  ``Evidence for an excess of $\bar{B} \to D^{(*)} \tau^-\bar{\nu}_\tau$ decays,''
  Phys.\ Rev.\ Lett.\  {\bf 109} (2012) 101802
 (\textit{Preprint} hep-ex/1205.5442)


\bibitem{Adachi:2009qg}
  I.~Adachi {\it et al.}  [Belle Collaboration],
  ``Measurement of B $\rightarrow$ $D^{(*)} \tau \nu$ using full reconstruction tags,''
  (\textit{Preprint} hep-ex/0910.4301)

\bibitem{Bozek:2010xy}
  A.~Bozek {\it et al.}  [Belle Collaboration],
  ``Observation of $B^+$ $\rightarrow$ $\bar {D^*}^0 \tau^+ \nu_{\tau}$ and Evidence for $B^+ \rightarrow \bar D^0 \tau^+ \nu_{\tau}$ at Belle,''
  Phys.\ Rev.\ D {\bf 82} (2010) 072005
  (\textit{Preprint} hep-ex/1005.2302)

\bibitem{Jung:2010ik} 
  M.~Jung, A.~Pich and P.~Tuzon,
  ``Charged-Higgs phenomenology in the Aligned two-Higgs-doublet model,''
  JHEP {\bf 1011}, 003 (2010)
   (\textit{Preprint} hep-ph/1006.0470)

\bibitem{Celis:2012dk}
  A.~Celis, M.~Jung, X.~-Q.~Li and A.~Pich,
  ``Sensitivity to charged scalars in $B\to D^{(*)}\tau\nu_\tau$ and $B\to\tau\nu_\tau$ decays,''
  JHEP {\bf 1301} (2013) 054
  (\textit{Preprint} hep-ph/1210.8443)


\bibitem{Nierste:2008qe}
  U.~Nierste, S.~Trine and S.~Westhoff,
  ``Charged-Higgs effects in a new B $\rightarrow$ D tau nu differential decay distribution,''
  Phys.\ Rev.\ D {\bf 78} (2008) 015006
(\textit{Preprint} hep-ph/0801.4938)


\bibitem{Kamenik:2008tj}
  J.~F.~Kamenik and F.~Mescia,
  ``$B \rightarrow D \tau \nu$ Branching Ratios: Opportunity for Lattice QCD and Hadron Colliders,''
  Phys.\ Rev.\ D {\bf 78} (2008) 014003 (\textit{Preprint} hep-ph/0802.379)





\bibitem{Tanaka:2010se}
  M.~Tanaka and R.~Watanabe,
  ``Tau longitudinal polarization in B $\rightarrow$ D tau nu and its role in the search for charged Higgs boson,''
  Phys.\ Rev.\ D {\bf 82} (2010) 034027
  (\textit{Preprint} hep-ph/1005.4306)


\bibitem{Fajfer:2012vx}
  S.~Fajfer, J.~F.~Kamenik and I.~Nisandzic,
  ``On the $B \to D^* \tau \bar \nu_\tau$ Sensitivity to New Physics,''
  Phys.\ Rev.\ D {\bf 85} (2012) 094025 (\textit{Preprint} hep-ph/1203.2654)



\bibitem{Sakaki:2012ft}
  Y.~Sakaki and H.~Tanaka,
  ``Constraints of the Charged Scalar Effects Using the Forward-Backward Asymmetry on $B\to D^{(*)}\tau\bar{\nu_{\tau}}$,''
 (\textit{Preprint}  hep-ph/1205.4908)





\bibitem{Fajfer:2012jt}
  S.~Fajfer, J.~F.~Kamenik, I.~Nisandzic and J.~Zupan,
  ``Implications of lepton flavor universality violations in B decays,''
  Phys.\ Rev.\ Lett.\  {\bf 109} (2012)  161801 (\textit{Preprint} hep-ph/1206.1872)



\bibitem{Datta:2012qk}
  A.~Datta, M.~Duraisamy and D.~Ghosh,
  ``Diagnosing New Physics in $b \to c \, \tau \, \nu_\tau$ decays in the light of the recent BaBar result,''
  Phys.\ Rev.\ D {\bf 86} (2012) 034027 (\textit{Preprint} hep-ph/1206.3760)



\bibitem{Crivellin:2012ye}
  A.~Crivellin, C.~Greub and A.~Kokulu,
  ``Explaining $B\to D\tau\nu$, $B\to D^*\tau\nu$ and $B\to \tau\nu$ in a 2HDM of type III,''
  Phys.\ Rev.\ D {\bf 86} (2012) 054014 (\textit{Preprint} hep-ph/1206.2634)







\bibitem{Tanaka:2012nw}
  M.~Tanaka and R.~Watanabe,
  ``New physics in the weak interaction of $\bar B\to D^{(*)}\tau\bar\nu$,''
  (\textit{Preprint} hep-ph/1212.1878)



\bibitem{Searches:2001ac}
  [LEP Higgs Working Group for Higgs boson searches and ALEPH and DELPHI and L3 and OPAL Collaborations],
  ``Search for charged Higgs bosons: Preliminary combined results using LEP data collected at energies up to 209-GeV,''
  (\textit{Preprint} hep-ex/0107031).
  

\bibitem{Pich:2009sp}
  A.~Pich, P.~Tuz\'on,
  ``Yukawa Alignment in the Two-Higgs-Doublet Model,''
  Phys.\ Rev.\ D {\bf 80} (2009) 091702
  (\textit{Preprint}  hep-ph/0908.1554)



\bibitem{Branco:2011iw}
  For a review, see for example:  G.~C.~Branco, P.~M.~Ferreira, L.~Lavoura, M.~N.~Rebelo, M.~Sher, J.~P.~Silva,
  ``Theory and phenomenology of two-Higgs-doublet models,''
  Phys.\ Rept.\  {\bf 516} (2012) 1
  (\textit{Preprint}  hep-ph/1106.0034);
  J.~F.~Gunion, H.~E.~Haber, G.~L.~Kane, S.~Dawson,
  ``The Higgs Hunter's Guide,''
  Front.\ Phys.\  {\bf 80} (2000) 1.


\bibitem{Hagiwara:1989gza}
  K.~Hagiwara, A.~D.~Martin and M.~F.~Wade,
  ``HELICITY AMPLITUDE ANALYSIS OF B $\rightarrow$ $D^*$ lepton neutrino DECAYS,''
  Phys.\ Lett.\ B {\bf 228} (1989) 144,
  ``Exclusive Semileptonic B Meson Decays,''
  Nucl.\ Phys.\ B {\bf 327} (1989) 569.



\bibitem{Korner:1989qb}
  J.~G.~Korner and G.~A.~Schuler,
  ``Exclusive Semileptonic Heavy Meson Decays Including Lepton Mass Effects,''
  Z.\ Phys.\ C {\bf 46} (1990) 93,
  ``Exclusive Semileptonic Decays of Bottom Mesons in the Spectator Quark Model,''
  ibid.\ C {\bf 38} (1988) 511
  [Erratum-ibid.\ C {\bf 41} (1989) 690].


\bibitem{Tanaka:1994ay}
  M.~Tanaka,
  ``Charged Higgs effects on exclusive semitauonic $B$ decays,''
  Z.\ Phys.\ C {\bf 67} (1995) 321
  (\textit{Preprint} hep-ph/9411405)

\bibitem{Chen:2005gr}
  C.~-H.~Chen and C.~-Q.~Geng,
  ``Lepton angular asymmetries in semileptonic charmful B decays,''
  Phys.\ Rev.\ D {\bf 71} (2005) 077501
  (\textit{Preprint} hep-ph/0503123)

\bibitem{Chen:2006nua}
  C.~-H.~Chen and C.~-Q.~Geng,
  ``Charged Higgs on $B \rightarrow \tau \bar \nu_{\tau}$ and $B \rightarrow P(V) \ell \bar \nu_{\ell}$,''
  JHEP {\bf 0610} (2006) 053
  (\textit{Preprint} hep-ph/0608166)



\bibitem{12070994}
  C.~McNeile, C.~T.~H.~Davies, E.~Follana, K.~Hornbostel and G.~P.~Lepage,
  ``Heavy meson masses and decay constants from relativistic heavy quarks in full lattice QCD,''
  Phys.\ Rev.\ D {\bf 86} (2012) 074503
  (\textit{Preprint} hep-lat/1207.0994)




\bibitem{Becirevic:2012jf}
  D.~Becirevic, N.~Kosnik and A.~Tayduganov,
  ``$\bar B\to D\tau\bar \nu_\tau$ vs. $\bar B\to D\mu\bar \nu_\mu$,''
  Phys.\ Lett.\ B {\bf 716} (2012) 208
 (\textit{Preprint}  hep-ph/1206.4977)

\bibitem{Bailey:2012jg}
  J.~A.~Bailey, A.~Bazavov, C.~Bernard, C.~M.~Bouchard, C.~DeTar, D.~Du, A.~X.~El-Khadra and J.~Foley {\it et al.},
  ``Refining new-physics searches in $B \rightarrow D \tau \nu$ decay with lattice QCD,''
  Phys.\ Rev.\ Lett.\  {\bf 109} (2012) 071802
  (\textit{Preprint} hep-ph/1206.4992)

\bibitem{John}
M. John, ``Prospects on $B \rightarrow D^* \tau \nu_{\tau}$ at LHCb,'' talk at  Workshop on $B$ decay into $D^{**}$ and related issues (2012).




\end{thebibliography}
\end{document}